\documentclass[review,12pt]{elsarticle}

\usepackage{amssymb}
\usepackage{amsmath}
\usepackage{hyperref}
\usepackage{latexsym,xcolor,multicol,calligra}

\journal{Chaos, Solitons and Fractals}

\begin{document}

\begin{frontmatter}

\title{Markov property of the Super-MAG Auroral Electrojet Indices}

\author[inst1]{Simone Benella}

\affiliation[inst1]{organization={INAF-Istituto di Astrofisica e Planetologia Spaziali},
            addressline={Via del Fosso del Cavaliere, 100}, 
            city={Rome},
            postcode={00133},
            country={Italy}}

\author[inst1]{Giuseppe Consolini}
\author[inst2,inst1]{Mirko Stumpo}
\author[inst1]{Tommaso Alberti}
\author[inst3]{Jesper W. Gjerloev}

\affiliation[inst2]{organization={Department of Physics, University of Rome Tor Vergata},
            addressline={Via della Ricerca Scientifica, 1}, 
            city={Rome},
            postcode={00133},
            country={Italy}}
            
\affiliation[inst3]{organization={Johns Hopkins University, Applied Physics Laboratory},
            city={Laurel},
            postcode={20723},
            country={MD, USA}}

\begin{abstract}

The dynamics of the Earth's magnetosphere exhibits strongly fluctuating patterns as well as non-stationary and non-linear interactions, more pronounced during magnetospheric substorms and magnetic storms. This complex dynamics comprises both stochastic and deterministic features occurring at different time scales. 
Here we investigate the stochastic nature of the magnetospheric substorm dynamics by analysing the Markovian character of SuperMAG SME and SML geomagnetic indices. By performing the Chapman-Kolmogorov test, the SME/SML dynamics appears to satisfy the Markov condition at scales below 60 minutes. The Kramers-Moyal analysis instead highlights that a purely diffusive process is not representative of the magnetospheric dynamics, thus a model that includes both diffusion and Poisson-jump processes is used to reproduce the SME dynamical features at small scales. A discussion of the similarities and differences between this model and the SME properties is provided with a special emphasis on the metastability of the Earth’s magnetospheric dynamics. Finally, the relevance of our results in the framework of Space Weather is also addressed.
\end{abstract}


\begin{highlights}
\item The statistics of the SME index satisfies the Markov property below 60 minutes
\item The small scale magnetospheric dynamics is modeled in terms of jump-diffusion process
\item The SME non-Markovian nature can be related to the metastability of the magnetosphere
\end{highlights}


\end{frontmatter}

\section{Introduction} \label{sec:introduction}

The near-Earth electromagnetic environment belongs to the class of far-from-equilibrium systems. Since early 90s it was realized that the short-time scale response of the Earth's magnetosphere to changes in the solar wind displays chaotic and nonlinear features \cite{Tsurutani1990,Baker1990} that cannot be simply described in terms of an input-output linear system \cite{Klimas1996}. In the last two decades, a clear evidence supporting complex dynamics has been provided in terms of hierarchical self-organized structures and criticality over a very wide range of time scales \cite{Klimas1996, Consolini1996,Consolini2001,Uritsky2002, Chang2003}. 

The  dynamics of the geospace plasma environment is mainly controlled by the changes of the physical conditions of the interplanetary medium, i.e., the solar wind and the interplanetary magnetic field. These changes, driven by the solar activity, affect the current systems flowing in both magnetosphere and ionosphere in terms of intensity and plasma convection, by increasing the plasma transfer towards the Earth's magnetospheric cavity \cite{Kamide1990, Gonzalez1994, Lyon2000, Borovsky2018}. The main manifestation of this interaction between the solar wind and the Earth's magnetosphere is the occurrence of magnetospheric substorms and magnetic storms. 

The response of the geospace plasma environment to changes in the interplanetary medium and the dynamics of the magnetosphere-ionosphere system can be monitored by using different indices, mainly derived from the temporal variations of ground-based magnetic field records from geomagnetic observatories. Some of these indices, such as the Auroral Electrojet (AE) and $D_{st}$-type indices, are proxies of the dynamics of the currents flowing in the magnetosphere-ionosphere system \cite{Davis1966, Iyemori1990}. Among these geomagnetic indices the AE-indices, which are related to the occurrence of magnetospheric substorms, provide one of the most relevant proxies to investigate the magnetospheric dynamics with a special emphasis on the dynamics of the Earth's magnetospheric central plasma sheet (CPS). Indeed, the dynamics of the plasma in the near-Earth CPS is strictly related to the currents flowing in the auroral regions (the auroral electrojets). However, although AE-indices are one of the best proxies for the substorm dynamics, they still present some limitations in correctly estimating the auroral electrojet current intensity in the case of strong geomagnetic activity, when the auroral oval expands to lower latitudes.  Nowadays, an attempt to overcome these limitations has been made by the SuperMAG collaboration through a generalization of the high-latitude indices, SME, SML and SMU, based on a large number (about 300) of currently operating ground-based geomagnetic observatories \cite{gjerloev2012supermag}. This new set of indices seems to be capable of better monitoring the magnetospheric dynamics during substorms.

Since their introduction, the AE indices have been extensively used as a proxy of the global magnetospheric dynamics and, in particular, to model the magnetospheric dynamics during substorms \cite{watkins2005towards, consolini2005magnetic, pulkkinen2006role,rypdal2010stochastic,chang1998self,consolini2005complexity}. For instance, Pulkkinen et al. \cite{pulkkinen2006role} used the AE index in order to show that the Langevin dynamics is quite satisfactory in reproducing the main statistical features of the magnetospheric dynamics in the frequency range [0.07, 3] mHz. They indeed found evidence that the AE probability distribution function (PDF) can be estimated by solving the stationary Fokker-Planck (FP) equation and that the statistics of waiting times between subsequent bursts, as evaluated through their model, is consistent with observations. 

As a matter of fact, the modeling of a great variety of physical processes through FP equation frequently allows one to capture important properties of the dynamics. In such cases, the processes satisfy the Markov condition and then the estimation of the first two Kramers-Moyal (KM) coefficients, i.e. the \textit{drift} and \textit{diffusion} terms, represents a fundamental starting point for both statistical analysis and modeling. This has been pointed out in many fields such as turbulence, economy, finance, neuroscience, cardiology, surface science and so forth \cite[e.g., see][for a comprehensive review on this topic]{friedrich2011approaching}. From the theoretical point of view, the FP equation is based on the assumption that the dynamics of the system can be fully described as a drift-diffusion process and then all the KM coefficients of order $\geq3$ vanishes (see Section \ref{sec:methods}). However, such crucial requirement has never been verified in the case of AE index, or, more in general, in the case of the magnetospheric dynamics. Moreover, moving to frequencies below 0.07 mHz, many of the aforementioned models lose their validity since the Markov property is no longer fulfilled.
The aim of this work is to apply the general analysis of Markov property and KM coefficients in the case of the magnetospheric dynamics in order to characterize its statistical properties over a wide range of time scales as well as to test the validity and limitation of our approach. To this purpose, we use the SuperMAG indices SME and SML in our analysis, instead of the AE, AL indices. 

Our work is organized as follows. A complete description of the data is presented in Section \ref{sec:data} and a summary of the methods involved in the analysis is provided in Section \ref{sec:methods}. The analysis presented in this work is structured in two parts: the first is referred to the investigation of the Markov property of SME and SML dynamics, Section \ref{sec:markov}, and the second one is devoted to the modeling of the SME index at small-scales in order to study the time-scale dependence of the Markov property, Section \ref{sec:model}. Finally, discussion and conclusions are drawn in Section \ref{sec:conclusions}.

\section{Data} \label{sec:data}


%
\begin{figure}
   	\centering
   	\includegraphics[width=\textwidth]{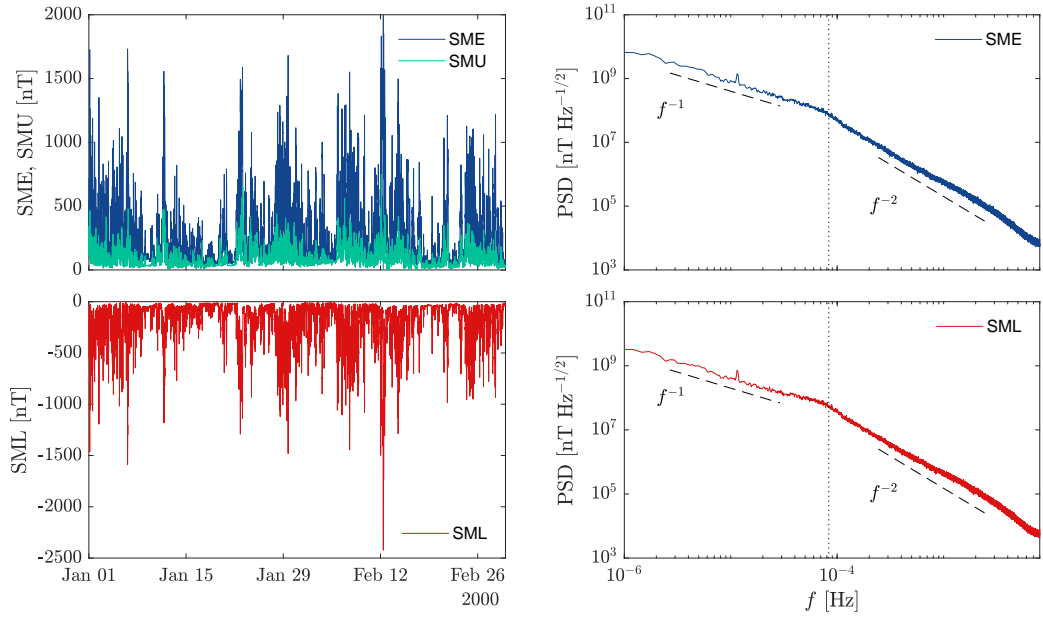}
   	\caption{Left: A sketch of the SME (blue line), SMU (green line) and SML (red line) indices timeseries during January and February 2000. Right: Power spectral density of the SME and SML indices. Dashed lines are a guide for the eye and represent the $f^{\beta}$ with $\beta=-1$ and $\beta=-2$ spectral trends. The vertical dotted line denotes the spectral break located at $\sim8\times10^{-5}$ Hz.}
   	\label{fig:data}
\end{figure}
In order to study the magnetospheric dynamics in response to interplanetary medium changes during substorms, in this work we use the Super-MAG auroral electroject indices SME and SML, which are a generalization of the traditional AE and AL indices, respectively \cite{newell2011evaluation}. These indices are based on a larger set of ground-based magnetic field measurements coming from about 300 geomagnetic observatories, thus covering a wider geographic area with respect to the smaller set of geomagnetic observatories used to compute AE-indices. For this reason, such larger dataset allows one to get a better estimation of the auroral electrojet current changes when the auroral oval extends to lower latitudes during strong geomagnetic substorms. 

The analysis is carried out by using the SME and SML indices timeseries at 1-minute resolution ranging from 1990 to 2015 and thus covering half of the Solar cycle 22, the entire cycle 23 and most of the cycle 24. Since these indices are rounded to integer nT values, we added a uniform white noise with strenght $\in[-0.5, 0.5]$ nT to the signal in order to avoid issues due to the digitalization. Figure \ref{fig:data} (left column) reports the timeseries of SME/SMU and SML during January and February 2000. We report only 2 months out of 25 years of data used in our analysis for visual purposes.
SME and SML power spectral densities calculated on the whole dataset are displayed in the right column of Figure \ref{fig:data}. Exactly as for the case of the AE and AL indices, they appear as colored noises $f^{-\beta}$ with two separate regimes: a $\beta\sim -1$ region below $\sim0.08$ mHz and a $\beta\sim -2$ region at higher frequencies. The spectral break of $\sim0.08$ mHz corresponds to $\sim200$ min and reflects some fundamental properties of the magnetospheric dynamics \cite{Tsurutani1990,Kamide1996,Consolini2005,newell2011substorm}. Indeed, temporal scales below 200 min have been associated with the occurrence of fast relaxation process occurring in the CPS of the magnetospheric tails, which are associated with the internal dynamics of the Earth's magnetosphere in response to interplanetary medium changes \cite{Klimas2000, Consolini2002, Uritsky2002}. In this work, we characterize and discuss this dynamical transition in terms of Markov processes.

\section{Methods} \label{sec:methods}

When dealing with empirical timeseries, the finite time resolution associated with measurements does not allow to adequately resolve the fast dynamics of an observed phenomenon. As a consequence, irregular high-frequency fluctuations are usually observed and can be assimilated to an overlaid noise term \cite{kantz2004nonlinear}. Starting from this observation we assume that the SME/SML timeseries can be described as a stochastic process $x(t)$ and we aim to test whether or not they satisfy the Markov condition. 

The Markov property is usually expressed in terms of the $n$-points transition probability of the process $x(t)$ as
\begin{equation}
    p[x_n,t+n\tau|x_{n-1}, t+(n-1)\tau; \dots; x_0, t]=p[x_n,t+n\tau|x_{n-1}, t+(n-1)\tau],
    \label{markov}
\end{equation}
where the transition probabilities can be defined through the Bayes' formula as
\begin{equation}
    p(x_1,t+\tau|x_0,t)=\frac{p(x_1,t+\tau;x_0,t)}{p(x_0,t)}.
    \label{bayes}
\end{equation}
Equations (\ref{markov}) and (\ref{bayes}) tell us that the knowledge of the initial distribution $p(x_0,t)$ and the transition probabilities allows to define the complete $n$-point probability distribution of the considered process. For a process satisfying the condition (\ref{markov}), the Chapman-Kolmogorov (CK) equation reads
\begin{equation}
    p(x_2,t+2\tau|x_0,t)=\int_{-\infty}^{+\infty}p(x_2,t+2\tau|x_1,t+\tau)p(x_1,t+\tau|x_0,t)dx_1.
    \label{eq:ck}
\end{equation}
The differential form of Equation (\ref{eq:ck}) expresses the time evolution of the transition probability and it is called \textit{master equation}, here reported in terms of the well-known Kramers-Moyal (KM) expansion \cite{risken1996fokker}
\begin{equation}
    \frac{\partial}{\partial t}p(x,t+\tau|x_0,t)=\mathcal{L}_{KM}(x)p(x,t+\tau|x_0,t),
    \label{kmexp}
\end{equation}
where $\mathcal{L}_{KM}(x)$ is the KM operator
\begin{equation}
    \mathcal{L}_{KM}(x)=\sum_{k=1}^\infty \biggl(-\frac{\partial}{\partial x} \biggr)^k D^{(k)}(x).
\end{equation}
The functions $D^{(k)}(x)$ are called KM coefficients and are defined by
\begin{equation}
    D^{(k)}(x)=\frac{1}{k!}\lim_{\tau\to0}\frac{M^{(k)}(x,\tau)}{\tau},
    \label{kmcoef}
\end{equation}
where $M^{(k)}$ are the conditional moments
\begin{equation}
    M^{(k)}(x,\tau)=\int_{-\infty}^{+\infty}(x'-x)p(x',t+\tau|x,t)dx'.
    \label{moments}
\end{equation}
As can be easily realized, the KM coefficients are not directly accessible from experimental timeseries, but rather one can compute the conditional moments at the data time resolution. 
Thus, it is possible to approximate the KM coefficients as their \textit{finite-scale} version evaluated for $\Delta\tau$ equals to the data time resolution $\tau_s$
\begin{equation}
    D^{(k)}_{\tau_s}(x)=\frac{1}{k!\tau_s}M^{(k)}(x,\tau_s).
    \label{eq:finitekm}
\end{equation}
Despite Equation (\ref{kmexp}) contains an infinite series of KM coefficients, the Pawula's theorem states that $D^{(k)}=0$ for $k\geq3$ if the fourth-order coefficient vanishes. In such case, the KM expansion reduces to the Fokker-Planck equation \cite{risken1996fokker}.

As previously mentioned, empirical timeseries may exhibit ﬂuctuations interrupted by sudden jumps between different states of the system, occurring in a very short time. However, such pronounced discontinuities can be introduced by the finite sampling of the underlying process that in principle can be or not a continuous diffusion process in the $\tau_s\to0$ limit. For a pure, statistically continuous, diffusive process, the state of the system as a function of time is described by the well-known Langevin equation
\begin{equation}
    dx_t=a(x)dt+b(x)dW_t,
    \label{eq:langevin}
\end{equation}
where $a(x)=D^{(1)}(x)$ is the \textit{drift} term and $b(x)=\sqrt{2D^{(2)}(x)}$ modulates the \textit{diffusion} term represented by the Wiener process $W_t$. However, the finiteness of the sampling time $\tau_s$ can introduce spurious effects in the \textit{finite-time} KM coefficient estimation such as non-vanishing high-order coefficients. A test for the Pawula's theorem which enables to state whether or not the process can be considered as a continuous diffusion, was instroduced by Lehnertz et al. \cite{lehnertz2018characterizing}. For a general Langevin process, the following linear relation between seocond- and fourth-order conditional moments for small values of $\tau$ exists \cite{lehnertz2018characterizing,tabar2019analysis}:
\begin{equation}
	M^{(4)}(x,\tau)\simeq 3[M^{(2)}(x,\tau)]^2.\label{eq:M2M4}
\end{equation}
The above relation can thus be used as a test of validity of the Pawula's theorem by inspecting only conditional moments which are directly accessible from data. In the following we will show that for the SME and SML indices the simple Langevin model (\ref{eq:langevin}) does not represent an accurate description of the observed dynamics and that the existence of non-vanishing high-order KM coefficients can be related to jump processes.

\section{Results}

\subsection{Analysis of the Markov property of SME and SML indices} \label{sec:markov}

We start our analysis of Markov features of the SME and SML indices by checking the validity of the CK equation at different time scales \cite{friedrich2011approaching,tabar2019analysis}. Let us refer to the left-hand side of Equation (\ref{eq:ck}) as \textit{empirical probability}, $p_E$, and to the right-hand side as \textit{CK probability}, $p_{CK}$. Equation (\ref{eq:ck}) states that $p_E=p_{CK}$ and the validity of such relation for the considered dataset provides information about the validity of the Markov property. We performed the CK test over different time scales and we report here three cases: $2\tau=10$ min, $2\tau=60$ min and $2\tau=200$ min. If we define the process $x_0(t)=x(t)$ as the SME index timeseries, the intermediate-scale process involved in the CK equation is $x_1(t,\tau)=x(t+\tau)$ and the large-scale process is defined as $x_2(t,\tau)=x(t+2\tau)$. We remark that the even spaced time separations are chosen for simplicity, nevertheless analogous results are obtained for any value of $0<\tau'<2\tau$ (not shown).

\begin{figure}
   	\centering
   	\includegraphics[width=\textwidth]{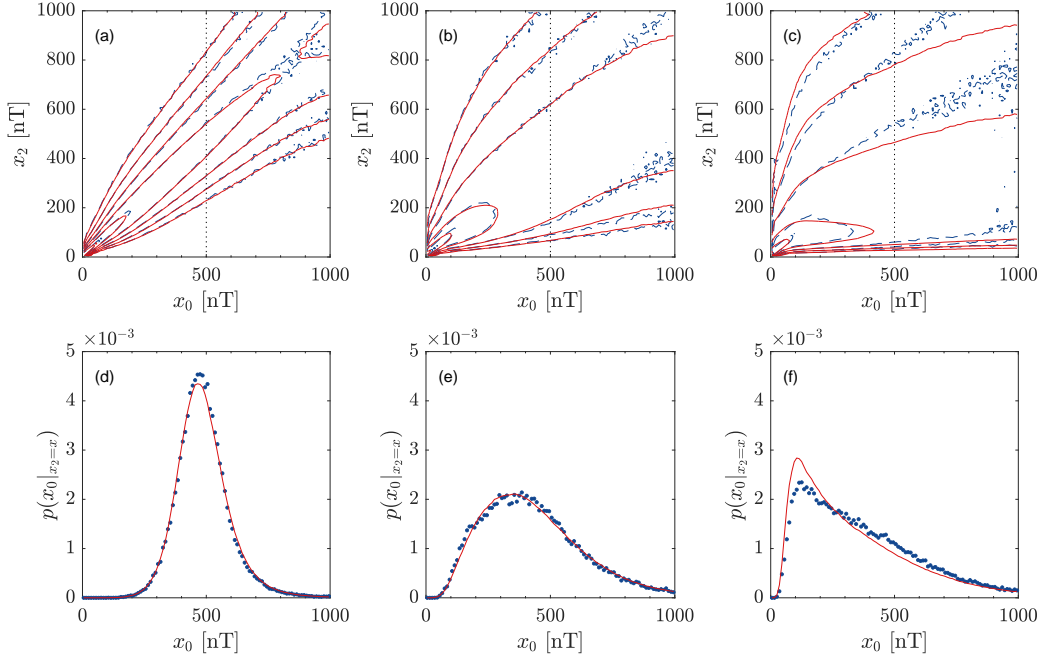}
   	\caption{Results of the CK test of the SME index at different time scales. In the top panels is reported the comparison between $p_E$ (blue) and $p_{CK}$ (red) for different time scales: (a) $2\tau=10$ min, (b) $2\tau=60$ min and (c) $2\tau=200$ min. In the bottom panels are reported PDFs obtained by evaluating the corresponding $p_E$ and $p_{CK}$ at $x_0=500$ nT for (d) $2\tau=10$ min, (e) $2\tau=60$ min and (f) $2\tau=200$ min.}
   	\label{fig:ck}
\end{figure}

The CK test results are reported in Figure \ref{fig:ck}. Figures \ref{fig:ck}a and \ref{fig:ck}b show a good agreement between the level curves obtained for $p_E$ (blue) and $p_{CK}$ (red), as also highlighted by evaluating the PDFs at $x_0=500$ nT (Figures \ref{fig:ck}d and \ref{fig:ck}e). Hence, we argue that the SME index satisfies the Markov property up to at least 60-minute time scales. Conversely, by looking at $2\tau = 200$ min (Figure \ref{fig:ck}c) we notice a clear deviation between the empirical distribution $p_E$ and the CK prediction $p_{CK}$ (see also the PDF at $x_0=500$ nT in Figure \ref{fig:ck}f). This finding suggests that at these scales the Markov condition is lost for the SME index. Thus, the fast dynamics of the SME index (i.e., at time scales shorter than 60 min) exhibits a ``memoryless'' character, whereas information contained in the history of the SME timeseries and/or in the external driver become important for the slow dynamics (i.e., $\tau \gtrsim60$ min). Similar results are also obtained for the SML index (not shown).

\begin{figure}
   	\centering
   	\includegraphics[width=\textwidth]{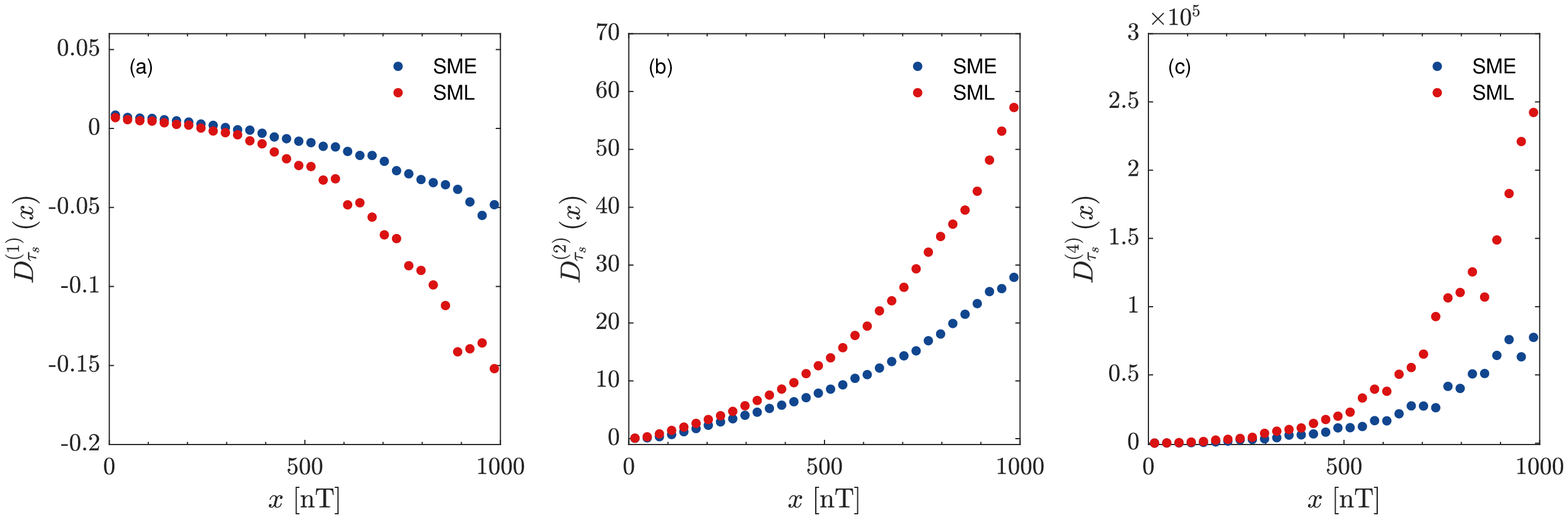}
   	\caption{Finite-time KM coefficients $D^{(1)}(x)$, $D^{(2)}(x)$ and $D^{(4)}(x)$ of the SME (blue) and $-\text{SML}$ (red) index time-series. The calculation is performed with $\tau=\tau_s=1$ min.}
   	\label{fig:km-coef}
\end{figure}

\begin{figure}
   	\centering
   	\includegraphics[width=.8\textwidth]{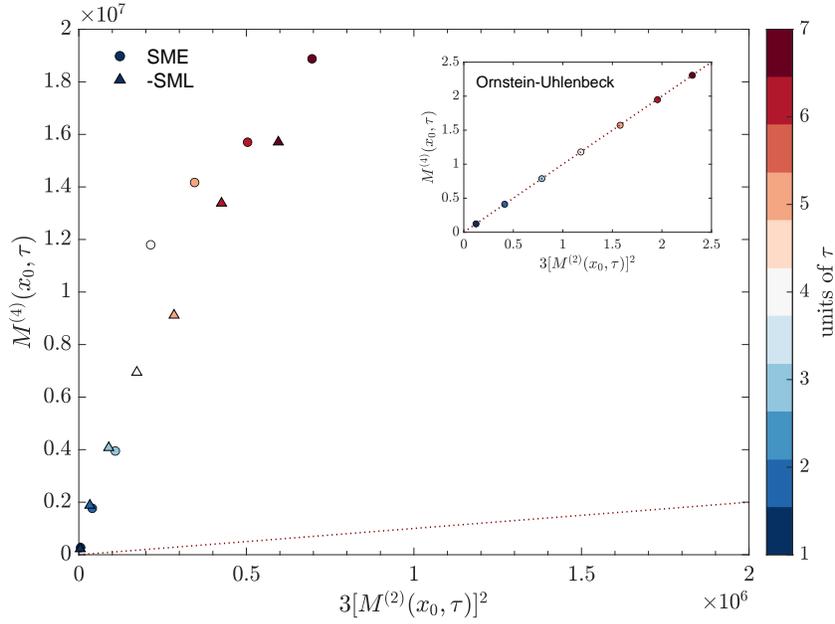}
   	\caption{Scatter plot of $M^{(4)}(x_0,\tau)$ vs $3[M^{(2)}(x_0,\tau)]^2$ for SME (circles) and $-\text{SML}$ (triangles) indices. The inset shows the same scatter-plot obtained in the case of the Ornstein-Uhlenbeck process. We chose $x_0$ around the peak of the probability distribution functions, i.e. $x_0\sim80$ nT for SME, $x_0\sim50$ nT for $-\text{SML}$ and $x_0\sim0$ for the Ornstein-Uhlenbeck process. The colormap indicates the different values of $\tau$ involved in the moment calculation in unit of time steps: $\tau=1$ min for SME/SML and $\tau=0.1$ for the integration time-step of the Ornstein-Uhlenbeck process. The red dotted line indicates the theoretical relation of Equation (\ref{eq:M2M4}) expected for a purely diffusive process.}
   	\label{fig:paw-test}
\end{figure}

Since SME and SML indices appear to be Markovian at small scales, we evaluate the first-, the second- and the fourth-order KM coefficients at the finite time scale $\tau_s=1$ min as a function of the index value (Figure \ref{fig:km-coef}). The interval for KM coefficients estimation considered here is limited to 1000 nT for both SME and $-\text{SML}$, since for larger values the statistics is not sufficient to safetely estimate the coefficients. First- and second-order KM coefficients, related to the \textit{drift} and \textit{diffusion} terms of the corresponding Langevin equation, are shown in Figures \ref{fig:km-coef}a, \ref{fig:km-coef}b and are consistent with previous findings based on the AE index analysis \cite{pulkkinen2006role}.
The same analysis, performed on the $-\text{SML}$ index, produces a similar trends of the KM coefficients although with different amplitudes. In particular, the deterministic term $D^{(1)}_{\tau_s}(x)$, associated with the drift, decreases more rapidly with respect to SME, whereas the stochastic term $D^{(2)}_{\tau_s}(x)$ shows a faster increase. A similar behavior is also observed for higher-order stochastic terms associated with $D^{(4)}_{\tau_s}(x,\tau)$, with both SME and SML indices having non-vanishing fourth-order KM coefficients (Figure \ref{fig:km-coef}c) that cannot be related to a finite-time effect. This would be expected since SME/SML increments involved in the estimation of the conditional moments of Equation (\ref{moments}) may assume quite large values and then a conditional average of the $k$-order power of such increments become progressively large for increasing values of $k$. By looking at the dependence of the moment $M^{(4)}_{\tau_s}(x,\tau)$ on $3[M^{(2)}_{\tau_s}(x,\tau)]^2$, Figure \ref{fig:paw-test}, a clear deviation from the expected theoretical linear behavior for a Langevin process (red line) is observed. For sake of comparison, we also report the same relation for an Ornstein-Uhlenbeck process with $D^{(1)}(x)=-x$ and $D^{(2)}(x)=1$, which satisfies Equation (\ref{eq:M2M4}) (see the inset of Figure \ref{fig:paw-test}). Hence, the non-vanishing high-order KM coefficients observed in this analysis constitute a physical feature of SME and SML index dynamics. The larger drift coefficient of SME with respect to SML for increasing values of $x$ suggests that the deterministic component of SME is mainly related to magnetospheric convective processes, described via SMU (we remark that $\text{SME}=\text{SMU}-\text{SML}$). Conversely, the larger values of high-order KM coefficients found for $-\text{SML}$ suggest that the major contribution to the stochastic terms of the SME dynamics can be ascribed to the burst-like activity of the geomagnetic tail.
\begin{figure}
   	\centering
   	\includegraphics[width=.8\textwidth]{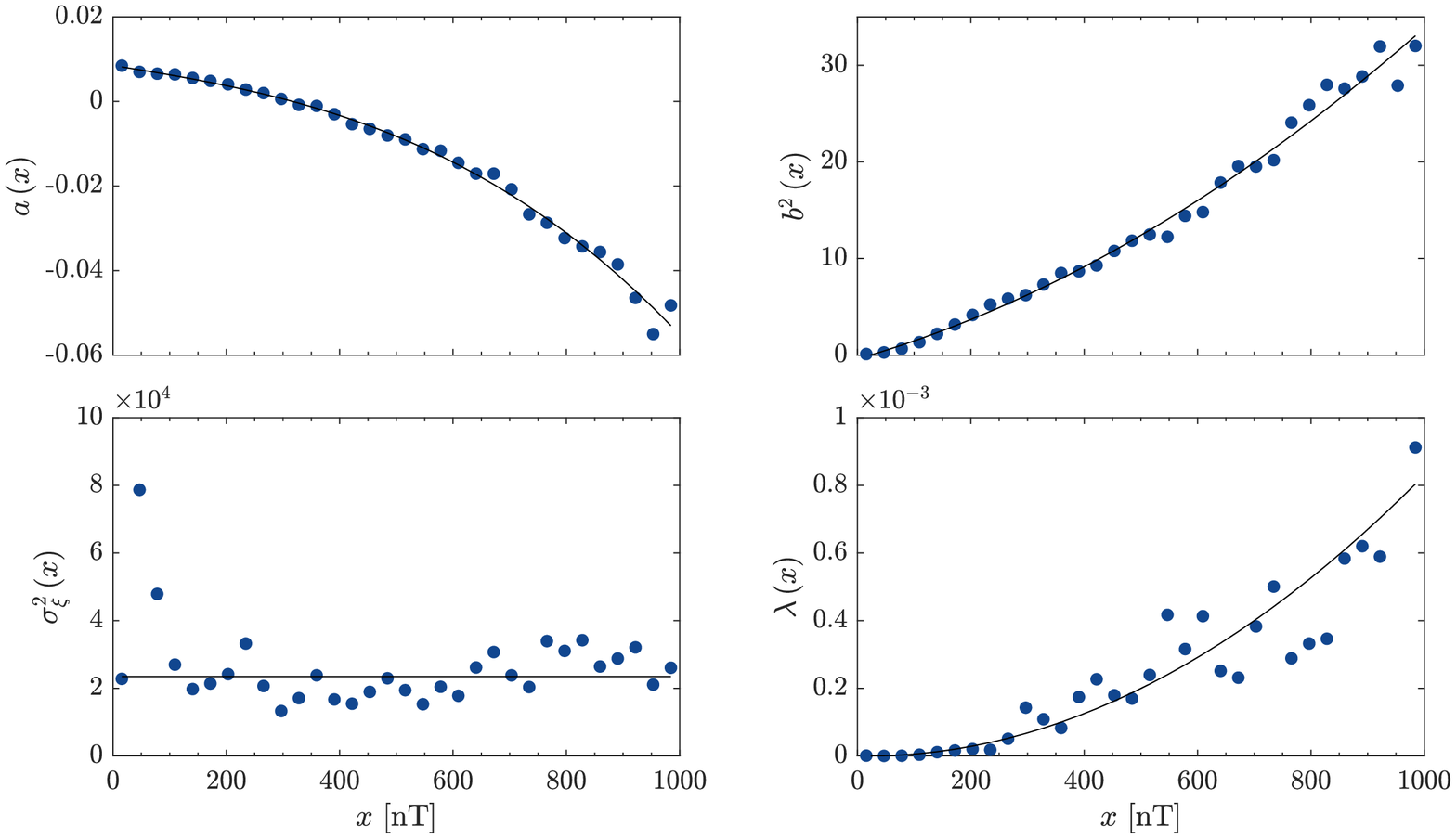}
   	\caption{Jump-diffusion process parameters estimated from the high-order conditional moments of the SME timeseries. The parameters are the drift term $a\,(x)$, diffusion term $b\,(x)$, jump amplitude $\sigma^2_\xi\,(x)$ and the jump rate $\lambda\,(x)$. Solid lines represent polynomial parameterizations.}
   	\label{fig:jumpdiff-param}
\end{figure}

\subsection{Jump-diffusion model for the SME index dynamics} \label{sec:model}

\begin{figure}
   	\centering
   	\includegraphics[width=8cm]{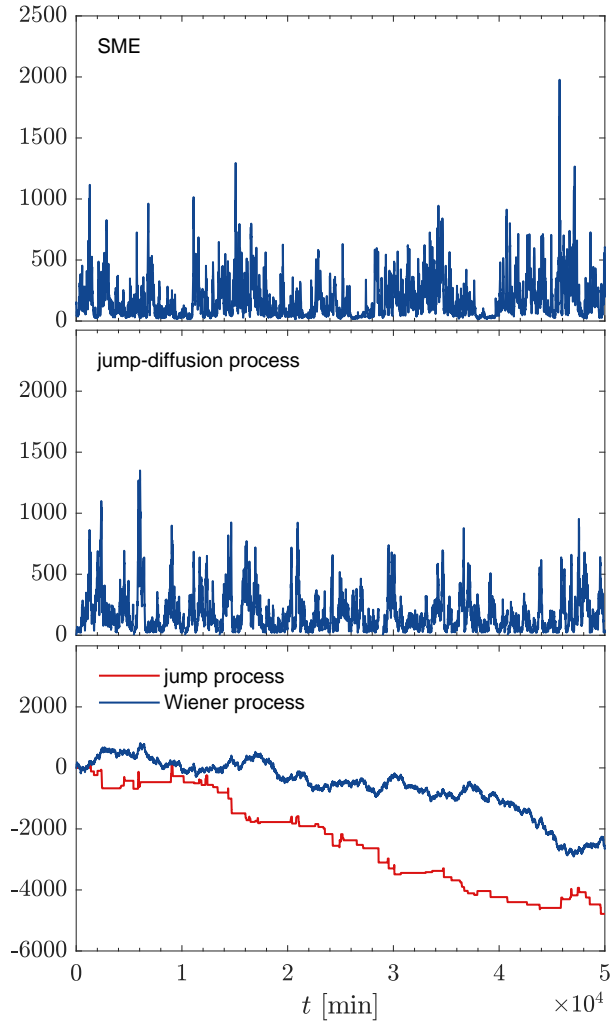}
   	\caption{Top panel: sketch of $10^4$ data points of the SME index timeseries. Middle panel: sketch of $10^4$ data points obtained by integrating the SDE model of Equation (\ref{eq:milst}) with the parameters of Figure \ref{fig:jumpdiff-param}. Bottom panel: diffusion and jump processes corresponding to the  timeseries depicted in the middle panel.}
   	\label{fig:sme-sde-comp}
\end{figure}

\begin{figure}
   	\centering
   	\includegraphics[width=10cm]{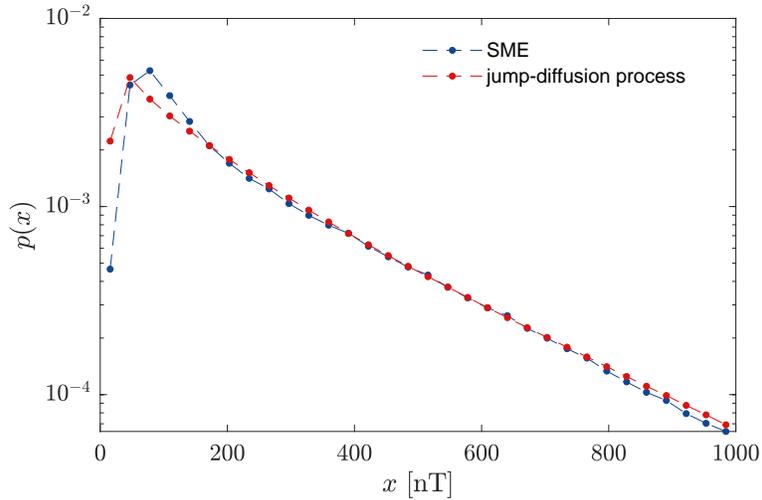}
   	\caption{Comparison between the PDFs of the SME index (blue dots) and the SDE model (red dots).}
   	\label{fig:sme-sde-pdf}
\end{figure}

As shown in the previous section, high-order KM coefficients are not vanishing for SME and SML indices. From a stochastic process perspective, this suggests that the magnetospheric dynamics cannot be represented as a continuous diffusion process, but it is characterized by jumps associated with the observed bust-like activity. Henceforth, we will focus on the SME index dynamics with the aim of showing that such index can be modeled as a stochastic process by adding a Poisson-like jump process to the Langevin dynamics \cite{tabar2019analysis}, i.e.,
\begin{equation}
     dx_t=a(x)dt+b(x)dW_t+\xi dJ_t,
     \label{eq:jumpdiff}
\end{equation}
where $a(x)$ is the drift term, $b(x)$ is the diffusion strength of the Wiener process $W_t$, and $\xi$ is the jump size of the Poisson jump process $J_t$. By assuming that the jump size $\xi$ follows a Gaussian distribution, i.e., $\xi\sim\mathcal{N}(0,\sigma_\xi)$, the terms appearing in Equation (\ref{eq:jumpdiff}) can be related to the KM coefficients \cite{anvari2016disentangling,tabar2019analysis}
\begin{eqnarray}
    a(x)&=&D^{(1)}(x), \nonumber \\
    b^2(x)&=&2D^{(2)}(x)-\lambda(x)\sigma^2_\xi(x), \label{eq:km-jumpdiff} \\
    \lambda(x)[\sigma^{2}_\xi(x)]^n &=& 2^n n! \, D^{(2n)}(x), \qquad n>2.\nonumber 
\end{eqnarray}
By inverting Equation (\ref{eq:km-jumpdiff}), we can estimate the jump size variance $\sigma^2_\xi(x)$, also called jump amplitude, and the jump rate $\lambda(x)$ as
\begin{equation}
    \sigma^2_\xi(x)=\frac{6 D^{(6)}(x)}{D^{(4)}(x)},\qquad     \lambda(x)=\frac{8D^{(4)}(x)}{\sigma^4_\xi}.
\end{equation}
The derived parameters of the jump-diffusion from the high-order KM coefficients are reported in Figure \ref{fig:jumpdiff-param}. The typical burst-like activity of the magnetospheric dynamics, joined with the lowering of the statistics for higher values of the SME index, produces a noisy estimation of the high-order conditional moments, affecting the estimation of jump parameters, as observed in Figure \ref{fig:jumpdiff-param}. 
In order to obtain the terms involved in the model, we parametrize amplitude and rate of the jump process directly from observations. In this case we do not expect any a priori functional dependence of such terms on the index values $x$ and we parametrize their trend by using polynomials as shown in Figure \ref{fig:jumpdiff-param} (black lines). The polynomial expressions for the jump-diffusion parameters associated with the SME index are
\begin{eqnarray*}
    a(x)&=&0.0085-2.08\times 10^{-5}x-7.60\times10^{-9}x^2-3.53\times10^{-11}x^3,\\
    b^2(x)&=&-0.40 + 0.017 x + 1.73\times10^{-5} x^2, \\
    \sigma_\xi^2(x)&=&2.35\times10^4, \\
    \lambda(x)&=&-3.22\times10^{-8}x+8.63\times10^{-10}x^2.
    \label{eq:jd-parameteriz}
\end{eqnarray*}
The drift  $a(x)$ and diffusion $b^2(x)$ terms are in agreement with previous findings by Pulkkinen et al. \cite{pulkkinen2006role} on the AE index, whereas the jump process appear to have a mean jump amplitude of $\sim2.35\times10^4$ nT$^2$ for any $x$ and an occurrence rate increasing as the SME index values with a quadratic trend.
The definition of the diffusion and jump processes allows us to integrate the Equation (\ref{eq:jumpdiff}) by using a jump-adapted strong integration scheme for stochastic differential equations (SDEs). Since the diffusion part is constituted by a multiplicative noise term, we adopt an integrator based on the Milstein scheme \cite{bruti2007approximation}
\begin{equation}
\begin{aligned}
\begin{split}
x_{t_{n+1-}}=x_{t_n}+a(x_{t_n})\Delta_{t_n}+b(x_{t_n})\Delta W_{t_n} +\\ \frac{b(x_{t_n})\,b'(x_{t_n})}{2}\{ (\Delta W_{t_n})^2 - \Delta_{t_n} \},
\end{split}\\
x_{t_{n+1}}=x_{t_{n+1-}}+c(x_{t_{n+1-}})\{ J_{t_{n+1}}-J_{t_{n+1-}} \}.
\end{aligned}\label{eq:milst}
\end{equation}
In Figure \ref{fig:sme-sde-comp} we report a comparison between 10$^4$ data points (for visual purposes) of the SME index (top panel) and a realization of the SDE integration (middle panel). According to \cite{pulkkinen2006role}, the integration time step $t_{n+1}-t_n$ is set equivalent to 5 s and the signal is later resampled to 1-minute resolution in order to be properly compared with SME index measurements. The Wiener and Poisson processes obtained through the stochastic integration on this particular sample are reported in the bottom panel. From a visual inspection we observe as many features of the SME index timeseries are present also in the jump-diffusion process. Middle and bottom panels show how the typical bursty character of the SME index can be well reproduced through the jump process. The model gives amplitudes of the bursts that are comparable in size with respect to those observed in the SME measurements and it also reproduces the SME nonzero baseline level. These features are also highlighted by the comparison between the PDFs of the SME index and the jump-diffusion process, Figure \ref{fig:sme-sde-pdf}. The comparison between the PDFs is shown in the interval $[0, 1000]$ nT, which is the interval over which diffusion and jump parameters are estimated. However, good agreement between PDFs extends up to $\sim2000$ nT, after which the SME statistics shows a sudden increase related to extreme events, sometimes called \textit{supersubstorms} \cite{tsurutani2015extremely}, which are not reproduced by the SDE.
\begin{figure}
   	\centering
   	\includegraphics[width=10cm]{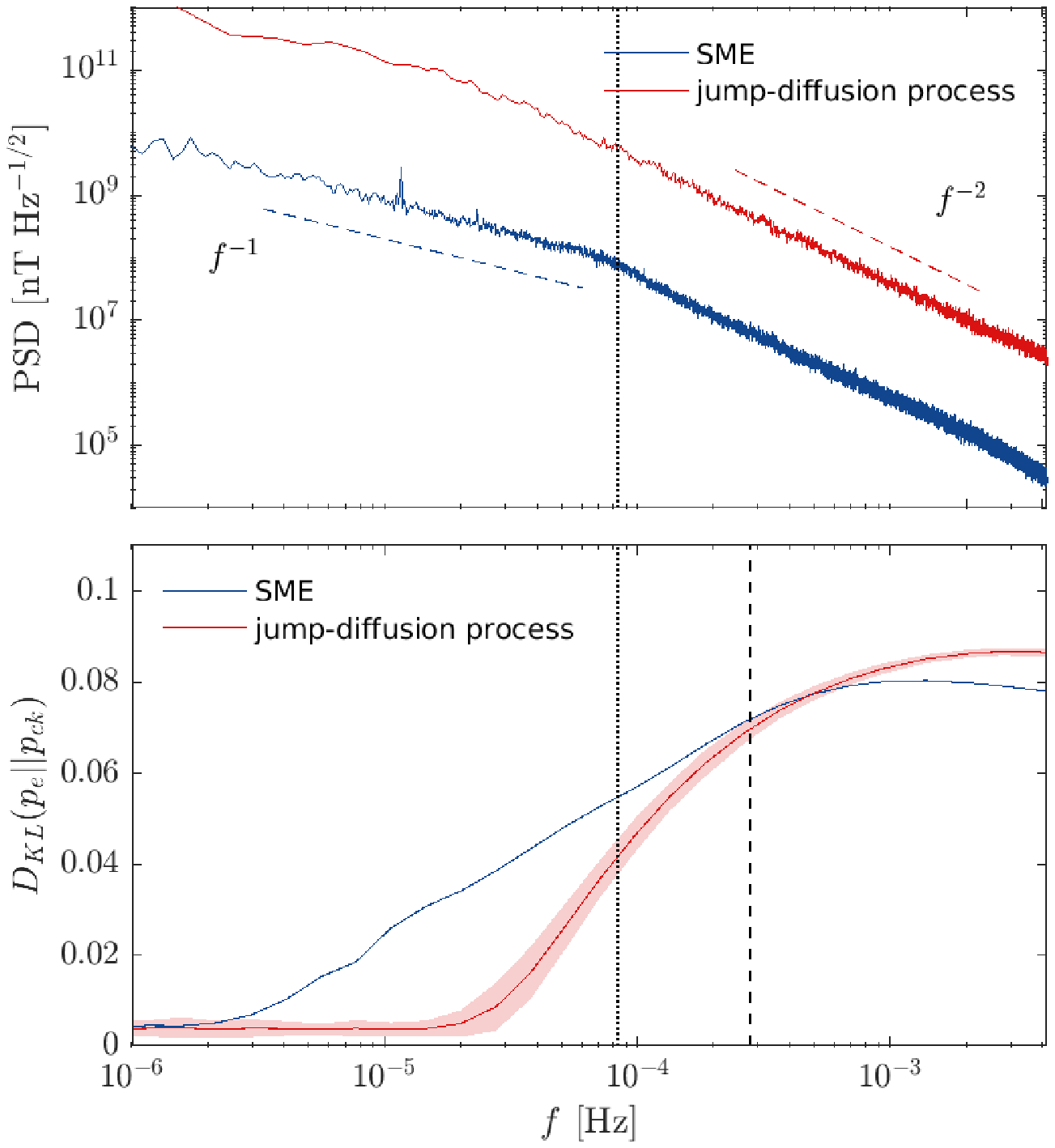}
   	\caption{Top panel: Power spectral density of the SME index (blue) and of the SDE model (red). The $f^{-1}$ and $f^{-2}$ trends are reported as a guide for the eye. The vertical dotted line indicates the spectral break of the SME index located at $\sim0.08$ mHz ($\sim200$ min). Bottom panel: KL divergence between $p_E$ and $p_{CK}$ for the SME index (blue) and the SDE model (red). In the case of the SDE model the curve is averaged over 20 independent realizations of the process and the red shaded area corresponds to the $\pm3\sigma$ bounds. The vertical dashed line indicates the frequency $\sim0.28$ mHz (corresponding to $\sim60$ min) at which the KL divergence of SME overcome the $3\sigma$ level of the model.}
   	\label{fig:kldiv}
\end{figure}
%

The CK test pointed out that the SME index statistics appears to violate the Markov property for time scales $\tau\sim200$ min (see Figure \ref{fig:ck}), whereas the model defined by Equation (\ref{eq:milst}) is constructed to satisfy the Markov property at any scale. Hence, it is natural to quantify the transition from a Markovian to a non-Markovian dynamics of the SME index by comparing its behavior to the one of a set of realizations of the model (Equation \ref{eq:milst}) as a function of the time scale. For this purpose, we introduce the Kullback-Leibler (KL) divergence as a functional enabling to quantify the goodness of the CK transition probability in approximating the empirical PDF as a function of the time scale $\tau$
\begin{multline}
    D_{KL}(p_E||p_{CK})=\iint p_{E}(x_2,t+2\tau;x_0,t)\times\\ \log\frac{p_{E}(x_2,t+2\tau;x_0,t)}{p_{CK}(x_2,t+2\tau;x_0,t)}dx_0\,dx_2.
\end{multline}
In Figure \ref{fig:kldiv} (top panel) we report the power spectral density (PSD) of the SME timeseries (black) along with the PSD of the timeseries obtained through the model (red). The comparison shows that unlike the SME index, the PSD associated with the model outcome exhibits the unique spectral trend $f^{-2}$. 
A comparison of the $D_{KL}(p_E||p_{CK})$ between SME (blue) and the model (red) as a function of the time scale is reported in the bottom panel. In the case of the SDE, we report the values of the KL divergence averaged over 20 realizations along with the $\pm3\sigma$ level (shaded area). We observe that the SME index can be described as a Markov process at small scales. Around $\tau\sim60$ min (vertical dashed line) the KL divergence associated with the SME index overcome the $3\sigma$ threshold of the $D_{KL}(p_E||p_{CK})$ associated with the model and at time scales above the spectral break, i.e. $\gtrsim200$ min, the deviation between the two curves becomes more pronounced. This result points out how the stochastic model can accurately capture the burst-like behavior of the magnetospheric dynamics, but also provides a quantitative information about the violation of the Markov property as a function of $\tau$. It is evident that for $\tau>60$ min, a deviation between the SME index statistics and the statistics expected for a Markov process begins. For increasing values of $\tau$ this discrepancy becomes more important suggesting that for a broad interval of scales, i.e. from 60 min up to $\sim10^4$ min, information contained in the history of the index itself and/or in the external driver become important and the Markov property is no longer satisfied by the magnetospheric activity. 

In this analysis we provide a characterization of the transition from Markovian to non-Markovian dynamics across a broad range of time scales in terms of the transition probabilities of the process pointing out as the stochastic process description is useful in unveiling different key properties of the complex magnetospheric dynamics.

\section{Discussions and conclusions}\label{sec:conclusions}

The high-latitude magnetospheric dynamics exhibits a very complex behavior. In this work we show that it is possible to disentangle the deterministic and stochastic parts constituting the overall SME and SML indices dynamics by applying a data analysis technique based on Markov process theory. The results confirm that the magnetospheric dynamics can be described as a Markov process at short time scales (i.e. $\lesssim60$ min) and that both SME and SML index dynamics present a burst-like activity that produces non-vanishing high-order KM coefficients. These coefficients show an increasing trend as a function of the index values and we showed that their contribution to the overall dynamics can be interpreted in terms of a Poisson-like jump process. We provide evidence that the simple framework of the jump process with Gaussian distributed amplitudes is quite accurate and capture many features of the SME dynamics, as the burst-like magnetospheric activity mainly associated with the occurrence of substorms. This jump dynamics is quite well consistent with the idea that the Earth's magnetospheric response during substorms as monitored by AE indices could resemble a fractional truncated L\'evy motion \cite{watkins2005towards}. Furthermore, the jump dynamics has to be  related to the occurrence of impulsive energy releases in the CPS region due to dynamical transitions in a complex fitness space \cite{Sitnov2001, Consolini2002b}. On the other hand, the Markov character of the dynamics of these indices at time scales shorter than 60 min could be the counterpart of the random superposition of sporadic intermittent reconnection and plasma energization/acceleration phenomena occurring in the geomagnetic CPS near-Earth region as observed in several works \cite{Lui1998, Angelopoulos1999, Consolini2002}. These sporadic events of energy relaxation appear as a random sequence of bursts in the AE index that can also partially superimpose one to the other without any characteristic time scale. As a consequence of these random temporal distribution of single activity bursts a $\sim 1/f^2$ interval can be observed in the power spectral density of these indices. This point of view is supported by simple numerical simulations \cite{Hwa1992,Klimas2000}.

An interesting result of our analysis is that we can associate the different terms (i.e., drift, diffusion and jump parameters) to different physical mechanisms. The SML dynamics, mainly related to the geomagnetic tail, contributes mostly in the burst-like behavior and can be represented as a stochastic process. Conversely, the global convective processes in the magnetosphere, mainly represented by the SMU index, result in a more deterministic contribution to the SME dynamics.

As a last step, we stress that the SDE enabling us to accurately model the SME dynamics, provides also a threshold for the Markov property of the real SME index. This is calculated as the KL divergence between the observed $p_E$ and the transition probability expected from the CK equation, $p_{CK}$. We find that the $3\sigma$ level with respect to the Markov condition are exceeded around 60 min. Hence, 
for time scales $\gtrsim60$ min, the history of the SME index starts to influence the dynamics in a non-trivial way and consequently
the deviation from the Markov property increases as a function of $\tau$.
In other words, there is a broad interval of scales where
the magnetospheric dynamics cannot be reproduced by assuming a simple ``memoryless'' process. At this long time scales the dynamics of the magnetosphere as monitored by SME displays a power spectral density mostly characterized by the $1/f$ domain discussed in several past works \cite{Consolini2002, Consolini2002b}, which could be associated with overlapping/interaction between single substorm bursts \cite{Hwa1992}.

Our results present fundamental implication in the field of Space Weather since our approach, based on data analysis in the framework of stochastic processes, is able to unveil some of the key properties of the complex magnetospheric activity in response to solar wind changes. Indeed, as also reported in a recent work using a stochastic approach \cite{Alberti18}, the overall dynamics of the magnetosphere, as represented by the low-latitude geomagnetic index SYM-H, consists of meta-stable states that are mainly driven and connected to the solar wind and interplanetary medium variability. By looking at the scale-dependence of the nature of these states Alberti et al. \cite{Alberti18} noted that the slow dynamics, occurring at scales larger than 200 min, is the main responsible of the observed meta-stability of the magnetospheric dynamics; conversely, the fast dynamics ($\tau \lesssim 100-200$ min) mainly persists in its stable state characterized by an increased level of stochasticity during geomagnetic storms, mainly related to the internal dynamics of the magnetosphere \cite{Alberti17}. Our results thus suggest that similar findings can be also observed for the high-latitude activity where a clear separation between directly-driven processes and loading/unloading mechanisms are present. The observed scale-separation highlighted in the present study in terms of the Markov property can be related to the different origin of processes involved into the overall dynamics of the SME/SML indices. While the ``memoryless'' character of the fast dynamics is mainly related to the occurrence of internal dynamical processes, only triggered by the solar activity, the non-Markovian nature that we found at long time scales can be associated with the driving effects of the surrounding solar wind. Thus, a proper modeling of the high-latitude variability must consider both drift and jump-diffusion processes to reproduce the burst-like component due to the geomagnetic tail activity and the forcing effects from the external solar wind. However, being the geospace environment a complex system, its components can react differently to the external forcing from the solar wind, thus producing different effects, both in terms of amplitude and occurrence, at both high and low latitudes. Our approach can be helpful to provide a simple model able to reproduce the statistical features of the high-latitude activity and can be used to provide thresholds in terms of auroral activity that can be used for Space Weather purposes. \\

%
%
The results presented in this paper rely on data collected at SuperMAG. We gratefully acknowledge the SuperMAG collaborators (\url{https://supermag.jhuapl.edu}). S.B. and G.C. acknowledge the financial support by Italian MIUR-PRIN grant 2017APKP7T on Circumterrestrial Environment: Impact of Sun-Earth Interaction. M.S. acknowledges the National Institute for Astrophysics, the University of Rome “Tor Vergata" and the University of Rome “La Sapienza" for the joint Ph.D. program “Astronomy, Astrophysics and Space Science". G.C., T.A. and J.W.G. acknowledge fruitful discussions within the scope of the International Team “Complex Systems Perspectives Pertaining to the Research of the Near-Earth Electromagnetic Environment” at the International Space Science Institute in Bern, Switzerland.

\bibliographystyle{elsarticle-num} 
\bibliography{markov_sme}

\end{document}